\documentclass[%
 reprint,
 amsmath,amssymb,
pra,
]{revtex4-2}

\usepackage{graphicx}
\usepackage{epstopdf}
\usepackage{dcolumn}
\usepackage{bm}
\usepackage{physics}
\usepackage{enumitem}
\usepackage{xcolor}

\begin{document}

\preprint{APS/123-QED}

\title{Nonlocality of the energy density for all single-photon states}

\author{M. Federico and H. R. Jauslin}
\email{jauslin@u-bourgogne.fr}
\affiliation{Laboratoire Interdisciplinaire Carnot de Bourgogne UMR 6303, CNRS, Universit\'e de Bourgogne, BP 47870, 21078 Dijon, France}%
\date{\today}%

\begin{abstract}
The nonlocality of single-photon states has been analyzed from several different but interrelared perspectives. In this article, we propose a demonstration based on the electromagnetic energy density observable and on the anti-local property of the frequency operator $\Omega=c(-\Delta)^{1/2}$. The present proof is based on the standard quantization of the electromagnetic field, which can be formulated equivalently in the momentum representations or in the position representations of Landau and Peierls [Z. Phys. {\bf 62}, 188 (1930)] and of Bia{\l}ynicki-Birula [\textit{Progress in Optics}, edited by E. Wolf (Elsevier, Amsterdam, 1996)]. 
Our proof extends to all single-photon states the results of Bia{\l}ynicki-Birula, which were formulated for two particular classes of states, those involving a uniform localization [Phys. Rev. Lett. {\bf80}, 5247 (1998)] or alternatively states that are electrically or magnetically localized [Phys.Rev. A {\bf79}, 032112 (2009)]. Our approach is formulated in terms of Knight's definition of strict localization [J. Math. Phys. {\bf 2}, 459 (1961)], based on the comparison of expectation values of single-photon states of local observables with those of the vacuum.
\end{abstract}

\maketitle

\section{Introduction}

Localization of photons, together with the existence of a position operator and consequently of a position wave function for photons, has maintained a vivid debate in the physics community since the early days of quantum mechanics \cite{Landau1930,Newton1949,Knight1961,Licht1963,Licht1966,Hegerfeldt1974,Hegerfeldt1980,Cook1982,BialynickiBirula1994,Sipe1995,BialynickiBirula1996,Hawton1999, Keller2005,Smith2007,Keller2016,Kiessling2018}. 
The question of the spatial localization of bosons in quantum field theory has followed several different but interrelated lines. 

One line is based on a definition of strictly localized states formulated by Knight \cite{Knight1961}. He showed that for states composed of a finite number of quanta, a class of  space- and time-dependent correlation functions of the Klein-Gordon field cannot be zero anywhere. Only states involving the superposition of infinitely many quanta, e.g., coherent states, can be spatially localized. This result was reformulated and extended by De Bi\`evre \cite{DeBievre2006,Bievre2007} in terms of expectation values of Weyl unitary operators.  In these type of approaches the main property that leads to nonlocality is the antilocal property of the frequency operator \cite{Segal1965,Masuda1972,Murata1973}, defined as the unique positive self-adjoint operator such that its square is minus the Laplacian: $\Omega= c (-\Delta)^{1/2}$. Antilocality means that if for some square-integrable field $\vec v(\vec x)$ both $\vec v=0$  and $\Omega \vec v =0$ in some finite volume, then $\vec v=0$ everywhere in ${\mathbb R}^3$.
In \cite{Licht1963,Licht1966}, following the results of Knight, Licht characterized the whole set of strictly localized states by showing that they can be obtained by applying a partial isometry $\hat W^\dag\hat W=1$ on the vacuum $\hat W^\dag\ket{\varnothing}$.

Bia{\l}ynicki-Birula showed that for two particular classes of single-photon states, the expectation value of the energy density cannot be zero in any finite volume.
The two classes considered are states with a spherically uniform support \cite{BialynickiBirula1998}, and states having either a magnetic or an electric localization property, defined in \cite{BialynickiBirula2009}. The argument in \cite{BialynickiBirula1998} was constructed using the Paley-Wiener theorem \cite{Paley1934}: Single photons with an exponential radial falloff of the form $\exp(-Ar)$, $A>0$, cannot exist but weaker falloffs are allowed. An example corresponding to a falloff of the form $\exp(-A\sqrt{r})$ was given. In general, in order to fulfill the constraint from the Paley-Wiener theorem, a quasiexponential localization is possible with a falloff $\exp(-Ar^\gamma)$, where $\gamma<1$. The advantage of the argument of \cite{BialynickiBirula1998} is that it provides a localization limit and gives a concrete example of a solution of Maxwell's equations approaching that limit. In \cite{BialynickiBirula2009} the authors introduced the notions of electrically and magnetically localized states and using a proof of the nonlocality of the helicity operator $\Lambda$ (see Appendix \ref{operators}), they showed that electrically or magnetically localized states cannot be local if one uses the energy density observable. 
The interest of the formulation of localization in terms of the expectation value of the energy density is that it is an observable that can be measured in current quantum optics experiments, using e.g. superconducting nanowire detectors \cite{Buller2009}. We show in Appendix \ref{Knight th} that Knight's theorem does not imply the results on the energy density since it does not apply to equal-time correlation functions.
The goal of the present article is to extend the results of Bia{\l}ynicki-Birula \textit{et al.} to all single-photon states.

We also mention \cite{Saari2005,Saari2012} where it was shown that there exist cylindrical functions for which a Gaussian falloff is possible in the waist plane only, making the localization stronger than the exponential limit shown in \cite{BialynickiBirula1998}. More recently in \cite{Gulla2021,Ryen2022,Gulla2021a}, some classes of strictly localized states were constructed (that are not single-photon states) so that they approach single-photon states as close as possible. 

Following the experimental development of on-demand single-photon sources \cite{Kuhn2002,McKeever2004,Scheel2009,Eisaman2011,Wang2019,Sinha2019} one can ask how it is possible to produce single photons in a controlled way without contradicting causality, or how close experimentally produced photons can be to perfect single-photon states. Similar questions concerning spontaneous emission have been addressed in \cite{Debierre2015,Debierre2016}. 
A precise analysis of these experiments can provide a test of whether the nonlocality predicted by the standard quantization is actually realized in experiments. As stated, e.g., in \cite{Gulla2021} the answer can be expected to be that in fact, those sources produce states that contain an infinity of multi-photon components (similar but different from coherent sates) that yield local energy densities, thus avoiding any conflict with causality. The presence of multiphoton components could in principle be tested in experiments, although one can expect that the multi-photon components will be quite small and will depend on the type of source considered.

In a second line, Hegerfeldt \cite{Hegerfeldt1974,Hegerfeldt1980,Hegerfeldt1994,Hegerfeldt1997} established some general properties of  bosons: For states containing exclusively positive energies, if the state function has a finite support at a given time, the time evolution will spread it over all space at any later times (see also \cite{Karpov2000}). In a later work, concerning causality in Fermi's two-atom model \cite{Hegerfeldt1997}, he showed that for positive bounded observables $0\leq\hat{\mathcal{O}}\leq1$, the expectation value $\bra{\psi(t)}\hat{\mathcal{O}}\ket{\psi(t)}$ is either nonzero for almost all $t$ or identically zero for all $t$. This opens the question of compatibility with Einstein causality. Some recent works have explored the 
possibility of constructing  variations of quantum field theory involving negative energy states that could avoid Hegerfeldt's nonlocality \cite{Mostafazadeh2006,Babaei2017,Hodgson2022,Hawton2023}.

A link between the positivity of energies and the anti-locality of the frequency operator $\Omega$ can be described as follows: The frequency operator can be defined as the square root of minus the Laplacian because $-\Delta$ is a positive operator, which is the reason why the energies are all positive in the Coulomb gauge quantization of the electromagnetic field.

In the present article we consider the standard quantization of the electromagnetic field in Fock space with a positive-energy spectrum, which is most often formulated in a momentum representation \cite{Mandel1995,Garrison2008}. As it was shown in \cite{Federico2023}, it can be formulated equivalently in two position representations: the Landau-Peierls (LP) \cite{Landau1930} and the Bia{\l}ynicki-Birula (BB) \cite{BialynickiBirula1996} representations. The equivalence of these quantizations is formulated in terms of isomorphisms of the Hilbert spaces with their corresponding scalar products \cite{Federico2023}. An advantage of the BB representation is that it is Lorentz covariant and independent of a choice of gauge, since it is formulated directly in terms of the electric and the magnetic fields. All our results are formulated in the Schr\"odinger representation but they do not involve at all the time evolution. The nonlocality of the expectation value of the energy density is proven at any fixed time, for all single-photon states. It is not created by the time evolution. This is an important difference from the works of Knight and Hegerfeldt.\\

\begin{figure}
\includegraphics[width=\columnwidth]{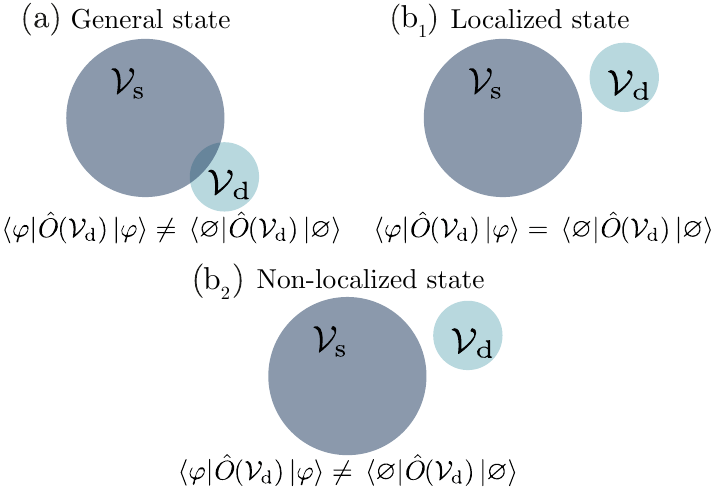}
\caption{Sketch illustrating Knight's definition of a localized state. We consider a state $\ket{\varphi}$ with an associated localization volume $\mathcal{V}_{\text{s}}$ and a detector $\hat O(\mathcal{V}_{\text{d}})$ with finite volume $\mathcal{V}_{\text{d}}$ which probes the state. Two situations can occur: (a) a general state where $\mathcal{V}_{\text{d}}\cap\mathcal{V}_{\text{s}}\neq\o$ and the expectation value of $\hat O(\mathcal{V}_{\text{d}})$ is not equal to that of the vacuum or (b$_1$) a localized state where $\mathcal{V}_{\text{d}}\cap\mathcal{V}_{\text{s}}=\o$ and the expectation value of $\hat O(\mathcal{V}_{\text{d}})$ is equal to that of the vacuum for all observables $\hat O$. This means that the state cannot be ``seen'' outside its volume of localization by any localized detector.
(b$_2$) A state is considered to be nonlocalized if it can be probed outside its apparent volume of localization by some observable $\hat O(\mathcal{V}_d)$, i.e., $\bra{\varphi}\hat O(\mathcal{V}_{\text{d}})\ket{\varphi}\neq\bra{\varnothing}\hat O(\mathcal{V}_{\text{d}})\ket{\varnothing}$ for $\mathcal{V}_{\text{d}}\cap\mathcal{V}_{\text{s}}=\o$. In general, this means that for such extended states, there is no $\mathcal{V}_{\text{s}}$ satisfying (a) and (b$_1$) for any $\hat O(\mathcal{V}_{\text{d}})$.}
\label{def-local-states}
\end{figure}

We recall the definition of a localized state that we will use as formulated by Knight \cite{Knight1961}: It is a state which cannot be detected by any means, outside its volume of localization, i.e., for any local observable, a localized state will give the same expectation value as the vacuum state, outside its volume of localization. For example, let $\ket{\varphi}$ be a spatially localized state in a volume $\mathcal{V}_{\text{s}}$ and $\hat O(\mathcal{V}_{\text{d}})$ a local observable which can probe the state $\ket{\varphi}$ over a volume $\mathcal{V}_{\text{d}}$. Since $\ket{\varphi}$ is localized, the expectation value of $\hat O(\mathcal{V}_{\text{d}})$ can give the following results [illustrated in FIG. \ref{def-local-states} (a) and (b$_1$)]:
\begin{itemize}
\item 
$\bra{\varphi}\hat O(\mathcal{V}_{\text{d}})\ket{\varphi}=\bra{\varnothing}\hat O(\mathcal{V}_{\text{d}})\ket{\varnothing}$ for $\mathcal{V}_{\text{d}}\cap\mathcal{V}_{\text{s}}=\o$,
\item
$\bra{\varphi}\hat O(\mathcal{V}_{\text{d}})\ket{\varphi}\neq\bra{\varnothing}\hat O(\mathcal{V}_{\text{d}})\ket{\varnothing}$ for $\mathcal{V}_{\text{d}}\cap\mathcal{V}_{\text{s}}\neq\o$.
\end{itemize}
The contraposition implies that a nonlocalized state is a state for which there exists at least one local observable whose expectation value is not equal to that of the vacuum at at least one point outside the volume of localization [FIG \ref{def-local-states} (b)].

In this article we will provide a demonstration of the nonlocalization of any single-photon state, based on the measurement of the mean value of the local energy density in any finite volume. It is inspired by the pioneering works of Bia{\l}ynicki-Birula and Bia{\l}ynicki-Birula \cite{BialynickiBirula1998,BialynickiBirula2009} and the antilocal property of the frequency operator $\Omega$ \cite{Segal1965,Masuda1972,Murata1973} (see Appendixes \ref{operators} and \ref{antilocal}). 

Another important operator that will be used in the following is the helicity operator $\Lambda$ \cite{BialynickiBirula2009,Federico2023}, defined by $c\nabla\times=\Omega\Lambda=\Lambda\Omega$ (see Appendix \ref{operators}). Its spectrum for transverse fields is $\{\pm1\}$ and it allows us to decompose any transverse field $\vec v$ into its positive and negative helicity parts $\vec v=\vec v^{(h+)}+\vec v^{(h-)}$ where $\vec v^{(h\pm)}=(1\pm\Lambda)\vec v/2$ and $\Lambda\vec v^{(h\pm)}=\pm\vec v^{(h\pm)}$.

The article is structured as follows. We first recall how one can describe single-photon states using position space representations. We then discuss the notion of a local observable and explain why the energy density is relevant in the context of local detection of single-photons. We finally formulate our proof of the nonlocality of photons and illustrate it with some examples.

\section{Representation of a general single-photon state}

In the quantum optics literature \cite{Mandel1995,Garrison2008,Fabre2020,Federico2022,Federico2023}, photons are often constructed using plane waves, which are not adapted to discuss their localization in position space. To overcome this issue, one can construct wavepackets from the general plane wave decomposition \cite{Mandel1995,Garrison2008,Fabre2020} or directly quantize the field using pulses of arbitrary shapes \cite{BialynickiBirula1996,BialynickiBirula1998,DeBievre2006,Bievre2007,BialynickiBirula2013,Federico2022,Federico2023,Fabre2023}.
In this article we will follow the latter choice, and taking advantage of the result shown in \cite{Federico2023}, we can switch back and forth between two equivalent position representations of the quantum theory. The first one, called the Landau-Peierls (LP) representation \cite{Landau1930,Federico2022,Federico2023}, is quite close to the standard plane waves quantization since it also takes the electric field and the vector potential as canonical variables, but in position space. The second one is what we call the Bia\l ynicki-Birula (BB) representation \cite{BialynickiBirula1996,BialynickiBirula1998,BialynickiBirula2013,Federico2023}, which takes as canonical variables the electric and magnetic fields.

The LP representation is based on the complex field
\begin{equation}
\vec\psi(\vec x)=\sqrt{\frac{\varepsilon_0}{2\hbar}}\left[\Omega^{1/2}\vec A(\vec x)-i\Omega^{-1/2}\vec E(\vec x)\right],\label{LP}
\end{equation}
where $\Omega^{\pm1/2}$ are self-adjoint operators constructed from the frequency operator introduced before (see Appendix \ref{operators}), $\vec A$ is the vector potential in the Coulomb gauge and $\vec x\in\mathbb{R}^3$. The LP field is an element of the Hilbert space of square-integrable functions from which we can construct a Fock space of states and annihilation and creation operators directly on arbitrary pulse-shaped functions \cite{Federico2022,Federico2023}
\begin{subequations}
\label{single photon state}
\begin{align}
\hat B_{\vec\psi}\ket{\varnothing}&=0,\\
\hat B^\dag_{\vec\psi}\ket{\varnothing}&=\ket*{\vec\psi},
\end{align}
\end{subequations}
satisfying the general bosonic commutation relations \cite{Garrison2008,Fabre2020}
\begin{equation}
\left[ \hat B_{\vec\psi},\hat B^\dag_{\vec\psi'} \right]=\bra*{\vec\psi}\ket*{\vec\psi'}_{LP}=\int_{\mathbb{R}^3}d^3x \ \vec\psi^\star\cdot\vec\psi'.\label{comm rel LP}
\end{equation}
The state $\ket*{\vec\psi}$ constructed in (\ref{single photon state}) represents a single-photon state carried by the classical pulse-shaped field $\vec\psi$. Its quantum dynamics is determined by the classical dynamics of the classical pulse \cite[Appendix B]{Federico2022} according to
\begin{equation}
\ket*{\vec\psi(t)}=\hat B^\dag_{\vec\psi(t)}\ket{\varnothing},
\end{equation}
where $\vec\psi(t)$ is a solution of the following complex representation of Maxwell's equations
\begin{subequations}
\begin{align}
i\frac{\partial\vec\psi}{\partial t}&=\Omega\vec\psi,\label{Max eq}\\
\nabla\cdot\vec\psi(t)&=0.\label{Max const}
\end{align}
\end{subequations}
This formulation of the quantum theory is related to the standard quantization using plane waves through the isomorphism which transforms $\vec\psi$ in the momentum space representation $z(\vec k,\sigma)$:
\begin{subequations}
\begin{align}
\vec\psi(\vec x)&=\int_{\mathbb{R}^3}d^3k\sum_{\sigma=\pm}\vec\phi_{\vec k,\sigma}(\vec x)z(\vec k,\sigma),\\
z(\vec k,\sigma)&=\int_{\mathbb{R}^3}d^3x\ \vec\phi_{\vec k,\sigma}^\star(\vec x)\cdot\vec\psi(\vec x),
\end{align}
\end{subequations}
where $\vec\phi_{\vec k,\sigma}$ are the plane waves of wave vector $\vec k$ and polarization $\sigma$ (see Appendix \ref{operators}). Creation and annihilation operators can thus be developed on the plane-wave basis 
\begin{subequations}
\begin{align}
\hat B^\dag_{\vec\psi}&=\int_{\mathbb{R}^3}d^3k\sum_{\sigma=\pm}z(\vec k,\sigma)\hat B^\dag_{\vec\phi_{\vec k,\sigma}},\\
\hat B_{\vec\psi}&=\int_{\mathbb{R}^3}d^3k\sum_{\sigma=\pm}z^\star(\vec k,\sigma)\hat B_{\vec\phi_{\vec k,\sigma}},
\end{align}
\end{subequations}
where $\hat B^\dag_{\vec\phi_{\vec k,\sigma}}$ and $\hat B_{\vec\phi_{\vec k,\sigma}}$ are the plane-wave creation and annihilation operators, respectively, in the position representation, i.e., the analog of the standard $\hat a^\dag_{\vec k,\sigma}$ and $\hat a_{\vec k,\sigma}$ in momentum space \cite{Mandel1995,Garrison2008,Fabre2020}. 

The BB representation can be constructed from the LP representation using the isomorphism $\mathcal{I}$ \cite{Federico2023} as
\begin{equation}
\label{isomorphism}
\vec F =\mathcal{I}\vec\psi=i\sqrt{\hbar}\Omega^{1/2}\vec\psi.
\end{equation}
Expressed in terms of the real electromagnetic variables, it takes the form
\begin{equation}
\vec F=\sqrt{\frac{\varepsilon_0}{2}}\left( \vec E+ic\Lambda\vec B \right),
\end{equation}
where we have used the relation $\Omega=c\Lambda\nabla\times$ (see Appendix \ref{operators}) and the definition of the vector potential $\nabla\times\vec A=\vec B$. The BB vector $\vec F$ is related to the Riemann-Silberstein vector $\vec F_{RS}=\sqrt{\varepsilon_0}(\vec E+ic\vec B)/\sqrt{2}$ through the helicity decomposition 
\begin{equation}
\vec F^{(h+)}=\vec F_{RS}^{(h+)},\quad\quad
\vec F^{(h-)}=\left(\vec F_{RS}^{(h-)}\right)^\star.
\end{equation}
The construction of creation and annihilation operators for the BB representation is directly given by the isomorphism $\mathcal{I}$
\begin{equation}
\hat C_{\vec F}=\mathcal{I}\hat B_{\vec\psi}\mathcal{I}^{-1},\quad\quad
\hat C_{\vec F}^\dag=\mathcal{I}\hat B^\dag_{\vec\psi}\mathcal{I}^{-1}.
\end{equation}
They satisfy the commutation relation 
\begin{equation}
\left[ \hat C_{\vec F},\hat C^\dag_{\vec F'} \right]=\bra*{\vec F}\ket*{\vec F'}_{BB}=\int_{\mathbb{R}^3}d^3x\ \vec F^\star\cdot\Omega^{-1}\vec F'.\label{comm rel BB}
\end{equation}
We have used another letter to refer to the creation and annihilation operators in the BB representation to emphasize that they do not act on the same Hilbert space \cite{Federico2023}, as one can see with the two different scalar products in the commutation relations (\ref{comm rel LP}) and (\ref{comm rel BB}), defining two Hilbert spaces 
\begin{subequations}
\begin{align}
\mathcal{H}_{LP}&=\left\{ \vec\psi\ \Big| \nabla\cdot\vec\psi=0, \bra*{\vec\psi}\ket*{\vec\psi'}_{LP}<\infty\right\},\\
\mathcal{H}_{BB}&=\left\{ \vec F\ \Big| \nabla\cdot\vec F=0, \bra*{\vec F}\ket*{\vec F'}_{BB}<\infty\right\}.\label{BB Hilbert space}
\end{align}
\end{subequations}
The BB scalar product (\ref{comm rel BB}) has the advantage that it is Lorentz invariant, and thus the BB representation is Lorentz covariant. Furthermore, it does not involve the choice of a gauge since the fields are expressed directly in terms of the electric and magnetic fields.

\section{Detection model --- Local energy observable}

We remark that in the quantum field theory of the electromagnetic field the localization properties cannot be established by looking only at the spatial properties of the state functions. One has to consider the joint representation of the states and the local observables. This can be done by considering e.g. mean values or correlation functions. The fact that the LP spatial properties do not correspond to the physically measurable properties had already been stated by Pauli \cite{Pauli1980}; we will give an explicit example of this particularity in Section \ref{illustration}. 

To show the nonlocality of single-photon states, it is enough to find one particular local observable for which Knight's localization criterion does not hold. An operator is said to be a local observable if it represents a physical measurement which can be made with an instrument well localized in space. Here $\vec{\hat E}(\vec x)$ and $\vec{\hat B}(\vec x)$ are local since in practice they can be measured by instruments involving e.g, localized charged particles or magnetic moments and thus possibly designed as small as required. Any operator that can be written as a point-wise function of $\vec{\hat E}(\vec x)$ and $\vec{\hat B}(\vec x)$ is also considered to be a local observable.

In this section, we introduce the local observable of the energy density that we will use to show the nonlocality, and which can represent some actual detectors. It was defined in \cite{BialynickiBirula1998} as
\begin{equation}
\hat{\mathcal{E}}_{\text{em}}(\vec x)=\frac{\varepsilon_0}{2}:\left(\vec{\hat E}^2(\vec x)+c^2\vec{\hat B}^2(\vec x)\right):,\label{local energy op}
\end{equation}
where $:\cdot :$ stands for the normal ordering. 

If one considers an experimental setup close to what is done in \cite{NisbetJones2011}, the photons that are produced are ``long'', i.e., carried by a pulse with a slowly varying envelope. Consequently, the associated pulse described in space, is much bigger than any actual detector. This means that the detector can probe the photon field only partially, without ``seeing'' the full state at the same time: $\mathcal{V}_{\text{s}}\gg\mathcal{V}_{\text{d}}$. Moreover, there exist efficient single-photon detectors, e.g., superconducting nanowires \cite{Buller2009}, that are sensitive to the electromagnetic energy. Those detectors are prepared at a temperature that is slightly below the critical tempeature $T_{\text{c}}$ of the superconducting nanowire, so it has no resistance. When a photon triggers the detector, there is a local absoption of energy, heating up the detector above $T_{\text{c}}$ and yielding a measurable resistance that signals the detection of a photon.

What is measured in such experiments, can thus be modeled by the local energy density $\hat{\mathcal{E}}_{\text{em}}(\vec x)$ integrated over the volume of the detector $\mathcal{V}_{\text{d}}$ i.e.
\begin{equation}
\hat{\mathcal{E}}_{\mathcal{V}_{\text{d}}}=\int_{\mathcal{V}_{\text{d}}} d^3x\ \hat{\mathcal{E}}_{\text{em}}(\vec x).
\end{equation}

The expectation value of the energy density operator for a general single-photon state in the LP representation, $\ket*{\text{1ph}}=\hat B_{\vec\psi}^\dag\ket{\varnothing}$ for any $\vec\psi\in\mathcal{H}_{LP}$ or equivalently written \cite{Federico2023} in the BB representation $\ket*{\text{1ph}}=\hat C_{\vec F}^\dag\ket{\varnothing}$ for $\vec F=i\sqrt{\hbar}\Omega^{1/2}\vec\psi\in\mathcal{H}_{BB}$, was computed in \cite{BialynickiBirula1998}. It can be written as (see Appendix \ref{details calcul} for an indication of the calculation)
\begin{subequations}
\label{mean value}
\begin{align}
\langle\hat{\mathcal{E}}_{\text{em}}(\vec x)\rangle_{\ket*{\text{1ph}}}&=\hbar\abs*{\Omega^{1/2}\vec \psi^{(h+)}(\vec x)}^2+\hbar\abs*{\Omega^{1/2}\vec \psi^{(h-)}(\vec x)}^2\\
&=\abs*{\vec F^{(h+)}(\vec x)}^2+\abs*{\vec F^{(h-)}(\vec x)}^2\geq0.\label{mean value BB}
\end{align}
\end{subequations}
We will show in the next section that this result implies the nonlocality of single photons.

\section{Proof of the nonlocality of the energy density for all single-photon states}\label{proof}

The result obtained above for the mean value of the energy density operator is clearly greater than or equal to zero for any single-photon pulse, i.e., for any function $\vec\psi$ or $\vec F$ representing the single-photon state. In this section we will show that the average local energy is strictly different from zero at any position in space 
\begin{equation}
\langle\hat{\mathcal{E}}_{\text{em}}(\vec x)\rangle_{\ket*{\text{1ph}}}\neq0\quad \forall \vec x\in\mathbb{R}^3.\label{main result}
\end{equation}
This result is a consequence of the following lemmas:
\\ \\
{\it
Lemma 1: For any field $\vec v(\vec x)$ that is not identically zero, $\Omega\vec v$ and $\vec v$ cannot be both zero in any open set of $\mathbb{R}^3$ 
}\cite{Segal1965,Masuda1972,Murata1973}.
\\ \\
{\it
Lemma 2: Fields of $\pm$ helicity, i.e. $\Lambda\vec g^{(h\lambda)}(\vec x)=\lambda\vec g^{(h\lambda)}(\vec x)$, $\lambda=\pm$, have the property that either $\vec g^{(h\lambda)}$ is identically zero or $\vec g^{(h\lambda)}\neq0$ in any open set of $\mathbb{R}^3$.}
\\ \\
A proof of {\it Lemma 1} is given in Appendix \ref{antilocal} and {\it Lemma 2} can be shown directly as follows: Let $\vec g$ be a transverse field, $\vec g=\vec g^{(h+)}+\vec g^{(h-)}$. The helicity components are eigenfunctions of the helicity operator $\Lambda\vec g^{(h\pm)}=\pm\vec g^{(h\pm)}$. Using the definition of the helicity operator $\Lambda=c\Omega^{-1}\nabla\times$, we can reformulate this relation as
\begin{equation}
c\nabla\times\vec g^{(h\pm)}=\pm\Omega\vec g^{(h\pm)}.\label{contradiction}
\end{equation}
Taking a given $\lambda=\pm$, if $\vec g^{(h\lambda)}$ is zero in an open set $\mathcal{S}$, then $\nabla\times\vec g^{(h\lambda)}$ is zero in the same set and finally (\ref{contradiction}) implies that $\Omega\vec g^{(h\lambda)}$ is also zero in $\mathcal{S}$, which by {\it Lemma 1} implies that $\vec g^{(h\lambda)}$ is zero everywhere. We conclude that $\vec g^{(h\lambda)}$ is either identically zero or non-zero for any open set as stated in {\it Lemma 2}.

The central result (\ref{main result}) can thus be shown by taking $\vec g$ to be the BB representation $\vec F$ of any single-photon state, which is transverse, and thus can be decomposed as $\vec F=\vec F^{(h+)}+\vec F^{(h-)}$. We conclude then that $\vec F^{(h\pm)}$ is either identically zero or nonzero in any open set. Moreover, since (\ref{mean value BB}) has two terms, even if one of them is identically zero, the other cannot be zero too since it would mean that the single-photon state itself is zero. This result is valid for any open set and therefore one can extend it to any point $\vec x\in\mathbb{R}^3$, which completes the proof.

Thus, since the zero point energy has been removed using the normal ordering in (\ref{local energy op}), $\langle \hat{\mathcal{E}}_{\text{em}}(\vec x) \rangle_{\ket*{1\text{ph}}}$ is never equal to the vacuum mean value, preventing Knight's localization criterion to be fulfilled for any single-photon state.
In physical terms this means that if the electromagnetic field is prepared in a single-photon state, a detector, placed anywhere in space, which measures the energy in a finite volume, has a non-zero probability of detecting the photon. The probability can be small, but it is strictly non-zero.

\section{Illustration of the nonlocality}
\label{illustration}

\begin{figure}
\includegraphics[width=\columnwidth]{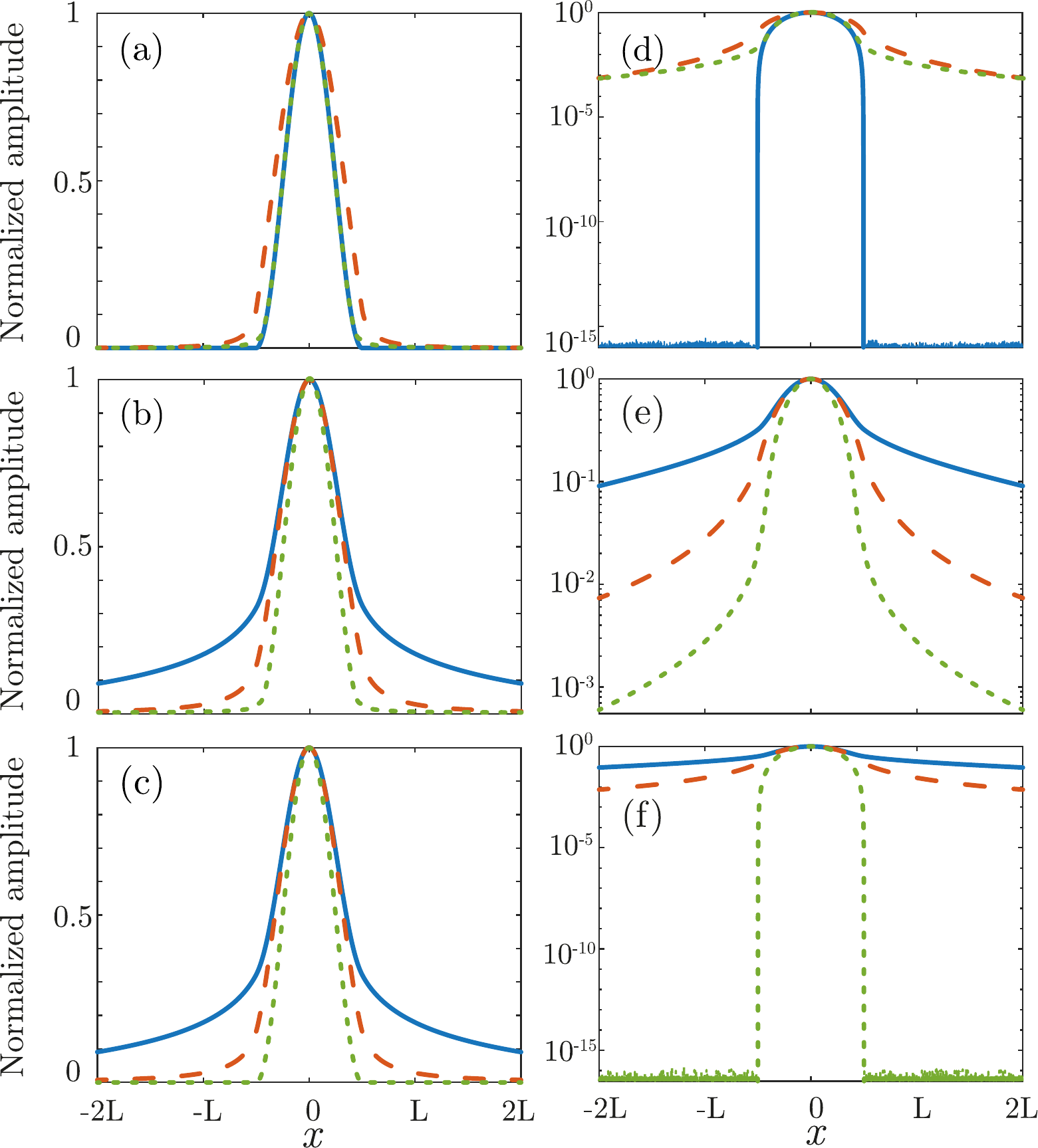}
\caption{Illustration of the nonlocality of single-photon states for two extreme cases. (a) The blue solid line shows the LP representation of a single-photon state $\ket*{\psi^{LP}_{\text{comp}}}$ with compact support and the green dotted line its BB representation. The red dashed line shows the expectation value of the energy density computed for that state. The compact support property of $\ket*{\psi^{LP}_{\text{comp}}}$ is lost for both the BB representation and the energy. (b) The blue solid line shows the LP representation of an extended single-photon state $\ket{\psi_{\text{ext}}}$ and the dotted green line its BB representation. The red dashed line shows the expectation value of the energy density computed from that state. The extended property of $\ket{\psi_{\text{ext}}}$ is visible from both representation and for the energy. (c) The green dotted line shows the BB representation of a single-photon state $\ket{F^{BB}_{\text{comp}}}$ with compact support and the solid blue line its LP representation. The red dashed line shows the expectation value of the energy density computed for that state. The compact support property of $\ket{F^{BB}_{\text{comp}}}$ is lost for the LP representation and the energy. (d), (e), and (f) are the same plots as (a), (b), and (c), respectively, but with a logarithmic scale.}
\label{nonlocal-energy-illus} 
\end{figure}

The nonlocality brought by the splitting into helicity components can be illustrated through simple one-dimensional examples. We can compute the expectation value $\langle\hat{\mathcal{E}}_{\text{em}}(\vec x)\rangle_{\ket{\text{1ph}}}$ for single-photon states representing three extreme cases: First we consider a state $\ket*{\psi^{LP}_{\text{comp}}}=\hat B_{\psi^{LP}_{\text{comp}}}^\dag\ket{\varnothing}$, where $\psi^{LP}_{\text{comp}}\in\mathcal{H}_{LP}$ is a function of compact support i.e. $\psi^{LP}_{\text{comp}}(x)=0$ outside an interval of size $L$; then we consider a state $\ket*{F^{BB}_{\text{comp}}}=\hat B_{F^{BB}_{\text{comp}}}^\dag\ket{\varnothing}$, where $F^{BB}_{\text{comp}}\in\mathcal{H}_{BB}$ is a function of compact support, i.e., $F^{BB}_{\text{comp}}(x)=0$ outside an interval of size $L$; and finally we consider a state $\ket*{\psi_{\text{ext}}}=\hat B_{\psi_{\text{ext}}}^\dag\ket{\varnothing}$, where $\psi_{\text{ext}}\in\mathcal{H}_{LP}$ is extended over all space, i.e., $\psi_{\text{ext}}(x)\neq0$ for any $x\in\mathbb{R}$. To construct $\psi_{\text{ext}}$, we use real fields $E_\text{comp}(x)$ and $A_{\text{comp}}(x)$ with support in the interval $[-L/2,L/2]$ of the form
\begin{subequations}
\begin{align}
E_{\text{comp}}(x)&\propto\left\{\begin{array}{ll}
\sin^2(\frac{\pi}{L}x+\frac{\pi}{2}) \quad \text{if $x\in[-\frac{L}{2},\frac{L}{2}]$}\\
0 \hspace{2.2cm} \text{otherwise},
\end{array}\right.\\
A_{\text{comp}}(x)&\propto\left\{\begin{array}{ll}
\sin^2(\frac{\pi}{L}x+\frac{\pi}{2}) \quad \text{if $x\in[-\frac{L}{2},\frac{L}{2}]$}\\
0 \hspace{2.2cm} \text{otherwise},
\end{array}\right.
\end{align}
\end{subequations}
and to build $\psi^{LP}_{\text{comp}}$ we take the extended fields
\begin{subequations}
\begin{align}
E^{LP}_{\text{ext}}(x)&=\Omega^{1/2}E_{\text{comp}}(x),\\
A^{LP}_{\text{ext}}(x)&=\Omega^{-1/2}A_{\text{comp}}(x).
\end{align}  
\end{subequations}
The resulting $\psi^{LP}_{\text{comp}}$ and $\psi_{\text{ext}}$ are represented as the solid blue lines in FIG. \ref{nonlocal-energy-illus}(a) and \ref{nonlocal-energy-illus}(d) and FIG. \ref{nonlocal-energy-illus}(b) and FIG. \ref{nonlocal-energy-illus}(e), respectively. To construct a localized BB representation $F^{BB}_{\text{comp}}$ we take the fields
\begin{subequations}
\begin{align}
E^{BB}(x)&=E_{\text{comp}}(x),\\
A^{BB}(x)&=\Omega^{-1}A_{\text{comp}}(x).
\end{align}
\end{subequations}
The resulting localized BB representation is shown by the green dotted line in FIG. \ref{nonlocal-energy-illus}(c) and \ref{nonlocal-energy-illus}(f). 

\noindent
For these three examples, we compute the expectation value of the energy density operator and obtain the results displayed as the red dashed lines in FIG. \ref{nonlocal-energy-illus}. In general, one can see that the localization property of the LP or BB representation does not give any information for the localization of the energy density. Indeed, the compact support property of the states $\psi^{LP}_{\text{comp}}$ and $F^{BB}_{\text{comp}}$ is not preserved for the expectation value of the local energy [FIG \ref{nonlocal-energy-illus}(a), \ref{nonlocal-energy-illus}(d), and FIG. \ref{nonlocal-energy-illus}(c), and \ref{nonlocal-energy-illus}(f)], as expected. Moreover, a localized LP representation implies a nonlocalized BB representation and vice versa due to the form of the isomorphism $\mathcal{I}$. This is illustrated in FIG \ref{nonlocal-energy-illus}(a), and \ref{nonlocal-energy-illus}(d) and FIG. \ref{nonlocal-energy-illus}(c), and \ref{nonlocal-energy-illus}(f) where we have either a localized LP function and a nonlocalized BB function or a nonlocalized LP function and a localized BB function. The most general case is shown in FIG. \ref{nonlocal-energy-illus}(b) and \ref{nonlocal-energy-illus}(e), where neither the LP nor the BB representation is localized and so neither is the energy density.

\section{Concluding remarks}

We point out that none of the statements in this article involve time evolution. They are statements about the expectation value of the energy density for any single photon state at any fixed time. This is an important difference from the works of Knight and of Hegerfeldt.
The nonlocality, as probed by the energy density, is not created by the time evolution; it is already present with the initial condition for any state as shown in FIG. \ref{nonlocal-energy-illus}.
Our point of view is that the nonlocality appears through the frequency operator $\Omega$ and it manifests in several ways: in time-independent expectation values like the energy density; in properties of the time evolution, as described by Hegerfeldt; or in time-dependent correlation functions, as treated by Knight. 
The time evolution of the states is determined by the frequency operator $\Omega$, both in the Landau-Peierls and in the Bia{\l}ynicki-Birula representations \cite{Federico2023}:
\begin{equation} 
i\frac{\partial \vec \psi}{\partial t} =  \Omega \vec \psi, \quad\quad\quad
i\frac{\partial \vec F}{\partial t} =  \Omega \vec F.
\end{equation}
The antilocality and the positivity of $\Omega$ are at the origin of the time-independent and the time-dependent nonlocality properties of single-photon states.
This different aspects are complementary and they can contribute to a better understanding of the nonlocality of single-photon states.

\section*{Acknowledgments}
We thank Jonas Lampart for many fruitful discussions. This work was supported by the ``Investissements d'Avenir'' program, project ISITE-BFC/IQUINS (ANR-15-IDEX-03), QUACO-PRC (ANR-17-CE40-0007-01) and the EUR-EIPHI Graduate School (17-EURE-0002).
We also acknowledge support from the European Union's Horizon 2020 research and innovation program under the Marie Sklodowska-Curie Grant Agreement No. 765075 (LIMQUET).

\begin{appendix}

\section{Plane wave basis, curl, frequency and helicity operators}\label{operators}

We consider the generalized basis of transverse plane waves $\{\vec\phi_{\vec k,\sigma}\}$ defined by
\begin{equation}
\vec\phi_{\vec k,\sigma}(\vec x)=(2\pi)^{-3/2}\vec\epsilon_\sigma(\vec k) e^{i\vec k\cdot\vec x},
\end{equation}
where $\sigma=\pm$, and the circular polarization vectors can be chosen as
\begin{equation}
\vec\epsilon_+(\vec k)=\frac{1}{\sqrt{2}|\vec k|\sqrt{k_x^2+k_y^2}}\left[
\begin{array}{c}
-k_xk_z+i|\vec k|k_y\\
-k_yk_z-i|\vec k|k_x\\
k_x^2+k_y^2
\end{array}\right],
\end{equation}
with $\vec\epsilon_-(\vec k)=\vec\epsilon_+(\vec k)^\star$. They are eigenfunctions of the curl operator with eigenvalues $\nabla\times\vec\phi_{\vec k,\sigma}=\sigma|\vec k|\, \vec\phi_{\vec k,\sigma}$. To construct the frequency operator $\Omega$, we recall that it is defined as the unique positive operator satisfying $\Omega^2=-c^2\Delta$ where $\Delta$ is the Laplace operator which can be written for transverse fields as $-\Delta=\nabla\times\nabla\times$. Therefore, the frequency operator satisfies $\Omega^2\vec\phi_{\vec k,\sigma}=c^2\abs*{\vec k}^2\vec\phi_{\vec k,\sigma}=\omega_{\vec k}^2\vec\phi_{\vec k,\sigma}$. The positive square root of $\Omega^2$ can thus be defined by its action on the continuum basis of plane waves
\begin{equation}
\Omega\vec\phi_{\vec k,\sigma}=\omega_{\vec k}\vec\phi_{\vec k,\sigma}, \:\:\:\:\: \omega_{\vec k}>0,
\end{equation}
and the particular powers $\Omega^{\pm1/2}$ used in the LP representation (\ref{LP}) can similarly be defined as $\Omega^{\pm1/2}\vec\phi_{\vec k,\sigma}=\omega_{\vec k}^{\pm1/2}\vec\phi_{\vec k,\sigma}$. Here $\Omega^2$, $\Omega$ and $\Omega^{\pm1/2}$ are all positive selfadjoint operators.

The helicity operator $\Lambda$ is then defined through a combination of the curl and the frequency operators by
\begin{equation}
c\nabla\times=\Omega\Lambda.\label{def Lambda}
\end{equation}
It has the same eigenfunctions with eigenvalues $\Lambda\vec\phi_{\vec k,\sigma}=\sigma\vec\phi_{\vec k,\sigma}$ and therefore commutes with both $\nabla\times$ and $\Omega$. One can decompose any transverse field $\vec v$ into a sum of a positive and a negative helicity part $\vec v^{(h\pm)}$
\begin{equation}
\vec v=\vec v^{(h+)}+\vec v^{(h-)},
\end{equation}
where $\Lambda\vec v^{(h\pm)}=\pm\vec v^{(h\pm)}$. The positive and negative helicity parts can be constructed by applying the projectors $\mathbb{P}^{(h\pm)}=(1\pm\Lambda)/2$, i.e. $\vec v^{(h\pm)}=\mathbb{P}^{(h\pm)}\vec v$. We also remark that $\Lambda^2=1$ and $\Lambda^{-1}=\Lambda$. 

Helicity can be interpreted as the projection of the spin on the direction of motion \cite{Federico2023}.

\section{A counterexample to the extension of Knigth's construction for equal-time correlation functions }\label{Knight th}

In this appendix we show that Knight’s result on time-dependent correlation functions cannot be extended to equal-time correlation functions, nor in particular to the expectation value of the energy density. Since we are interested in photons, we adapt Knight's construction for the scalar Klein-Gordon equation to the electromagnetic case but a similar argument can be made for massive scalar fields. In fact the argument is based on the construction that was made by Bia{\l}ynicki-Birula and Bia{\l}ynicki-Birula in \cite{BialynickiBirula2009} in order to show that there exist single-photon states with either electric or magnetic localization (but not both). 

We start with a state $\ket*{\vec\psi}=\hat B^\dag_{\vec\psi}\ket{\varnothing}$ with the properties 
\begin{subequations}
\label{local prop correlation}
\begin{align}
\left[\hat B_{\vec\psi},\hat B^\dag_{\vec\psi}\right]&=1,\\
\left[\vec{\hat A}(\vec x_j,t_0),\hat B^\dag_{\vec\psi}\right]&=0\quad\forall\vec x_j\notin\mathcal{R},\label{local prop correlation 2}
\end{align}
\end{subequations}
where $\mathcal{R}$ is an open set of $\mathbb{R}^3$ and $t_0$ a given time which we take as zero in the following without lost of generality.
The mean value of the equal-time correlation function of the potential vector outside $\mathcal{R}$ for a state with such properties is thus
\begin{align}
&\bra*{\vec\psi}\vec{\hat A}(\vec x_1)\dots\vec{\hat A}(\vec x_M)\ket*{\vec\psi}=\bra*{\varnothing}\hat B_{\vec\psi}\vec{\hat A}(\vec x_1)\dots\vec{\hat A}(\vec x_M)\hat B_{\vec\psi}^\dag\ket*{\varnothing}\nonumber\\
&\quad\quad\quad=\bra*{\varnothing}\hat B_{\vec\psi}\hat B_{\vec\psi}^\dag\vec{\hat A}(\vec x_1)\dots\vec{\hat A}(\vec x_M)\ket*{\varnothing}\nonumber\\
&\quad\quad\quad=\bra*{\varnothing}\left(1+\hat B^\dag_{\vec\psi}\hat B_{\vec\psi}\right)\vec{\hat A}(\vec x_1)\dots\vec{\hat A}(\vec x_M)\ket*{\varnothing}\nonumber\\
&\quad\quad\quad=\bra*{\varnothing}\vec{\hat A}(\vec x_1)\dots\vec{\hat A}(\vec x_M)\ket*{\varnothing}.
\end{align}
This calculation shows that outside $\mathcal{R}$, the mean value is that of the vacuum. The key point of this result is to construct a state $\ket*{\vec\psi}$ with the properties (\ref{local prop correlation}). To do so, the commutator (\ref{local prop correlation 2}) can be written as
\begin{align}
\left[\vec{\hat A}(\vec x_j),\hat B^\dag_{\vec\psi}\right]&=\sqrt{\frac{\hbar}{2\varepsilon_0}}\int_{\mathbb{R}^3}d^3k\sum_{\sigma=\pm}\omega_{\vec k}^{-1/2}\vec\phi_{\vec k,\sigma}(\vec x_j)\left[\hat B_{\vec\phi_{\vec k,\sigma}},\hat B^\dag_{\vec\psi}\right]\nonumber\\
&=\sqrt{\frac{\hbar}{2\varepsilon_0}}\Omega^{-1/2}\int_{\mathbb{R}^3}d^3k\sum_{\sigma=\pm}\vec\phi_{\vec k,\sigma}(\vec x_j)\bra*{\vec\phi_{\vec k,\sigma}}\ket*{\vec\psi}\nonumber\\
&=\sqrt{\frac{\hbar}{2\varepsilon_0}}\Omega^{-1/2}\vec\psi(\vec x_j),
\end{align}
which is zero outside $\mathcal{R}$ if one takes a function $\vec\xi$ with compact support in $\mathcal{R}$ i.e. $\vec\xi(\vec x_j)=0$ for all $\vec x_j\notin\mathcal{R}$ and defines $\vec\psi=\Omega^{1/2}\vec\xi$.

\section{Derivation of the mean value of the energy density for a general single-photon state}
\label{details calcul}

In order to have a self-contained proof of nonlocality discussed in this article, we provide here an indication of the main steps for calculating the mean value of the energy density given in (\ref{mean value}). For simplicity, we will show here only the derivation done with the BB representation of the states, but one can equivalently compute it with the LP representation. The result can anyway be expressed easily in both representations using the isomorphism $\mathcal{I}$ [see eq. (\ref{isomorphism})]. 

We first remark that the normal-ordered energy density observable (\ref{local energy op}) can be written as \cite{BialynickiBirula1998}
\begin{equation}
\hat{\mathcal{E}}_{\text{em}}=\frac{\varepsilon_0}{2} :\left(\vec{\hat E}-ic\vec{\hat B}\right)\cdot\left(\vec{\hat E}+ic\vec{\hat B}\right):,\label{em energy}
\end{equation}
and that electromagnetic field operators can be expressed as
\begin{subequations}
\label{elm fields op}
\begin{align}
\vec{\hat E}&=\frac{1}{\sqrt{2\varepsilon_0}}\left(\vec{\hat{{\bf F}}}+\vec{\hat{{\bf F}}}^\dag\right),\\
\vec{\hat B}&=\frac{-i}{\sqrt{2\varepsilon_0c^2}}\Lambda\left(\vec{\hat{{\bf F}}}-\vec{\hat{{\bf F}}}^\dag\right),
\end{align}
\end{subequations}
in terms of the BB field operator $\vec{\hat{{\bf F}}}$ defined by \cite{Federico2023}
\begin{equation}
\vec{\hat{{\bf F}}}(\vec x)=\int_{\mathbb{R}^3}d^3k\sum_{\sigma=\pm}\vec g_{\vec k,\sigma}(\vec x)\hat  C_{\vec g_{\vec k,\sigma}}.\label{BB field op1}
\end{equation}
The set of functions $\{\vec g_{\vec k,\sigma}\}$ introduced here is a basis of the BB Hilbert space $\mathcal{H}_{BB}$ (see eq. (\ref{BB Hilbert space})). By inserting equations (\ref{elm fields op}) into (\ref{em energy}) and using the decomposition $\vec{\hat{{\bf F}}}=\vec{\hat{{\bf F}}}^{(h+)}+\vec{\hat{{\bf F}}}^{(h-)}$ which can be done with the projectors $\mathbb{P}^{(h\pm)}=(1\pm\Lambda)/2$, we can rewrite the energy density as
\begin{equation}
\hat{\mathcal{E}}_{\text{em}}(\vec x)=\ :\left(\vec{\hat{{\bf F}}}^{(h+)\dag}+\vec{\hat{{\bf F}}}^{(h-)}\right)\cdot\left(\vec{\hat{{\bf F}}}^{(h+)}+\vec{\hat{{\bf F}}}^{(h-)\dag}\right):.\label{energy density field op}
\end{equation}
Among the four terms of (\ref{energy density field op}), only two will give non zero expectation values and thus we have
\begin{equation}
\langle\hat{\mathcal{E}}_{\text{em}}\rangle_{\ket*{1 \text{ph}}}=\sum_{\lambda=\pm}\bra{\varnothing}\hat C_{\vec F}\vec{\hat{{\bf F}}}^{(h\lambda)\dag}\cdot \vec{\hat{{\bf F}}}^{(h\lambda)}\hat C_{\vec F}^\dag\ket{\varnothing}.
\end{equation}
To work out the final result, we use the commutators
\begin{equation}
\left[ \vec{\hat{{\bf F}}}^{(h\pm)}(\vec x),\hat C_{\vec F}^\dag \right]=\vec F^{(h\pm)}(\vec x),
\end{equation}
and obtain (\ref{mean value}) in the BB representation
\begin{equation}
\langle\hat{\mathcal{E}}_{\text{em}}\rangle_{\ket*{1 \text{ph}}}=\abs*{\vec F^{(h+)}}^2+\abs*{\vec F^{(h-)}}^2.
\end{equation}

\section{Anti-locality of the frequency operator}\label{antilocal}

We provide in this appendix a proof of {\it Lemma 2} used to show the nonlocality of photons in Section \ref{proof}. This result was shown in \cite{Segal1965,Masuda1972,Murata1973} and we will sketch here the argument of \cite{Masuda1972}. We recall the statement.
\\ \\
{\it
Lemma 2: For any field $\vec v(\vec x)$ that is not identically zero, $\Omega\vec v$ and $\vec v$ cannot be both zero in any open set of $\mathbb{R}^3$. 
}
\\ \\
We are going to prove that if $\vec v$ and $\Omega\vec v$ are both equal to zero in some open set $\mathcal{S}$, it imples that $\vec v(\vec x)=0$ everywhere.

\noindent
\textit{Proof}. Since $\Omega$ is positive and selfadjoint, the operators $U(t)=\exp(i\Omega t)$, $t\in(-\infty,+\infty)$ define a one-parameter family of unitary operators. The field defined as $\vec u(\vec x,t)=U(t)\vec v(\vec x)$ satisfies the wave equation
\begin{equation}
\frac{\partial^2\vec u}{\partial t^2}=-\Omega^2\vec u,
\end{equation}
with initial conditions
\begin{subequations}
\begin{align}
\vec u(\vec x,t=0)&=\vec v(\vec x),\\
\frac{\partial\vec u}{\partial t}(\vec x,t=0)&=i\Omega\vec v(\vec x).
\end{align}
\end{subequations}
Since the solutions of the wave equation propagate with a finite speed $c$, the property of the initial conditions to be zero i.e. $\vec v(\vec x)=0$ in a set $\mathcal{S}$, implies that there is a $t_0 > 0$ and a non-empty open subset $\mathcal{S}_0\subset\mathcal{S}$ such that $\vec u(\vec x,t)=0$ for all $\vec x\in\mathcal{S}_0$ and $0\leq t<t_0$. Thus, for any $\mathcal{C}^{\infty}$ field $\vec\varphi(\vec x)$ with compact support in $\mathcal{S}_0$
\begin{equation}
\bra*{\vec\varphi}\ket*{\vec u(\cdot,t)}=\int_{\mathbb{R}^3}d^3x\ \vec\varphi(\vec x)^\star\cdot\vec u(\vec x,t)=0
\end{equation}
for $0\leq t<t_0$. We now consider the continuation of the variable $t$ into the upper complex half plane and define the function
\begin{equation}
f(z)=\bra*{\vec\varphi}\ket{e^{i\Omega z}\vec v},
\end{equation}
for $\Im z\geq 0$ that has the following properties \cite{Masuda1972}:
\begin{enumerate}[label=(\roman*)]
\item $f(z)$ is holomorphic for $\Im z > 0$, and continuous for $\Im z \geq 0$,\label{prop 1}
\item $f(t) = \bra*{\vec\varphi}\ket*{\vec u(\cdot,t)}$ when $t \in\mathbb{R}$, and $f(t)\in\mathbb{R}$ for $t\in(0, t_0)$,\label{prop 2}
\item $f(t) = 0$ for $0 \leq t < t_0$.\label{prop 3}
\end{enumerate}
We remark that for $t > t_0$, $f(t)$ is not necessarily zero or real. The main steps of a proof that $f(z)$ is holomorphic in the upper half-plane, by showing the existence of the derivative $df/dz$, can be summarized as follows: $\vec v$ can be developed in the basis $\vec \phi_{\vec k,\sigma}$ of continuum eigenfunctions of the Laplacian $\vec v = \int_{\mathbb{R}^3} d^3k\sum_\sigma \alpha_{\vec k,\sigma}\vec \phi_{\vec k,\sigma}$.  We can then write
\begin{align}
\frac{d}{dz}f(z) &= \frac{d}{dz} \int_{\mathcal{S}_0} d^3x\ \vec \varphi^\star \cdot e^{i\Omega z}  \vec v \nonumber\\
&= \int_{\mathcal{S}_0} d^3x\ \vec \varphi^\star\cdot e^{i \Omega z}i\Omega \vec v \nonumber\\
&= \int_{\mathcal{S}_0} d^3x\int_{\mathbb{R}^3} d^3k\sum_\sigma \vec \varphi^\star \cdot \vec \phi_{\vec k,\sigma}  \alpha_{\vec k,\sigma} i \omega_{\vec k}e^{i \omega_{\vec k} z}. 
\end{align}
The derivative can be brought inside the integral by the Weierstrass M-test and the Lebesgue dominated convergence theorem, because the integral converges absolutely since $e^{i \Omega z} $ is a contractive semi-group \cite{Talvila2001}. The existence of the derivative is then obtained by exchanging the two integrals by the Fubini-Tonelli theorem: Writing $z$ as $z=z_r+iz_i$, one obtains 
\begin{align}
\abs{\frac{d}{dz}f(z)}&\leq  \int_{\mathcal{S}_0} d^3x\int_{\mathbb{R}^3} d^3k\sum_\sigma \abs*{\vec \varphi^\star \cdot \vec \phi_{\vec k,\sigma}} \abs*{\alpha_{\vec k,\sigma}} \omega_{\vec k}e^{- \omega_{\vec k} z_i}\nonumber\\
&=\int_{\mathbb{R}^3} d^3k\sum_\sigma \abs*{\alpha_{\vec k,\sigma}} \omega_{\vec k}e^{- \omega_{\vec k} z_i}\int_{\mathcal{S}_0} d^3x\abs*{\vec \varphi^\star \cdot \vec \phi_{\vec k,\sigma}} \nonumber\\
&\leq C\int_{\mathbb{R}^3} d^3k\sum_\sigma \abs*{\alpha_{\vec k,\sigma}} \omega_{\vec k}e^{- \omega_{\vec k} z_i}<\infty,
\end{align}
where $C$ is a constant. The last integral is finite for $z_i>0$ which completes the argument for the existence of $df/dz$.

We will now use the Schwarz reflection principle \cite[p.75]{Cartan1963}, which in the present context can be formulated as follows: If $f_+(z)$ satisfies the properties that 
\begin{enumerate}[label=(S-\roman*)]
\item $f_+(z)$ is holomorphic in the open upper complex rectangle $D_+ = \left\{\Im z > 0, \Re z \in (0, t_0)\right\}$,
\item $f_+(z)$ continuous in $D_+\cup (0, t_0)$,
\item $f_+(t)$ is real in the interval $t \in (0, t_0)$,
\end{enumerate}
then $f_+(z)$ can be continued holomorphically trough the interval $(0, t_0)$ to the lower rectangle $D_- = \left\{\Im z \leq 0,\Re z\in(0,t_0)\right\}$, by defining 
\begin{align}
f_+(z)=\left\{\begin{array}{ll}
f_+(z)\:\:\:\:\:\:\:\:\:\: \text{for $z\in D_+$},\\
(f_+(z^\star))^\star \:\: \text{for $z\in D_-$}.
\end{array}
\right.
\end{align}
This defines thus a holomorphic function $f_+(z)$ in the whole open set $D_+\cup D_- = \left\{\Re z \in (0, t_0)\right\}$, which includes the interval $(0, t_0)$.

By applying the Schwarz reflection principle to the function $f(z)$ that we combine with the properties \ref{prop 1} and \ref{prop 2}, it shows that $f(z)$ is analytic in the union of the open upper half-plane and $D_-$. Property \ref{prop 3} states that $f(z) = 0$ in the interval $z \in (0, t_0)$, which implies that $f(z) = 0$ in the whole region where $f$ is holomorphic, in particular in the whole open upper half-plane $\Im z > 0$. Since according to \ref{prop 1}, $f(z)$ is continuous for $\Im z \geq 0$, this implies
that $f(z)=0$ for $\Im z\geq0$. In particular, $f(t):\bra*{\vec\varphi}\ket*{\vec u(\cdot,t)}=0$ for $-\infty<t<\infty$. Since $\vec\varphi$ is an arbitrary function, this implies that $\vec u(\vec x,t)=0$ for $-\infty<t<\infty$ and $\vec x\in\mathcal{S}_0$. The unique continuation theorem for solutions $\vec u(\vec x,t)$ of the wave equation, proven e.g. in \cite{Masuda1967}, states that if $\vec u(\vec x,t)=0$ for an open set for all $-\infty<t<\infty$, then $\vec u(\vec x,t)=0$ everywhere. In particular, for $t=0$ this implies that
\begin{equation}
\vec u(\vec x,t=0)=\vec v(\vec x)=0, \forall\vec x\in\mathbb{R}^3,
\end{equation}
which completes the proof.

\end{appendix}

\bibliography{localization_biblio}

\end{document}